\documentclass[twocolumn,prl,aps,amsmath,amssymb,epsfig]{revtex4}

\usepackage{graphicx}% Include figure files
\usepackage{dcolumn}% Align table columns on decimal point
\usepackage{bm}% bold math
\usepackage{feynmp}

\newcommand{\bea}{\begin{eqnarray}}
\newcommand{\eea}{\end{eqnarray}}
\begin{document}
\title{Electroweak Symmetry Breaking via Bose-Einstein Mechanism}
 \author{Francesco {\sc Sannino}}
 \email{francesco.sannino@nbi.dk}
 \author{Kimmo {\sc Tuominen}}\email{tuominen@nordita.dk}
 \affiliation{{\rm NORDITA}, Blegdamsvej 17, DK-2100 Copenhagen \O, Denmark }
\date{May 2003}

\begin{abstract}
Recently the Bose-Einstein phenomenon has been proposed as
possible physical mechanism underlying the spontaneous symmetry
breaking in cold gauge theories. The mechanism is natural and we
use it to drive the electroweak symmetry breaking. It can be
implemented in different ways while here we present a simple model
in which the Bose-Einstein sector is felt only indirectly by all
of the standard model fields. The structure of the corrections due
to the new mechanism is general and independent on the model used
here leading to experimental signatures which can be disentangled
from other known extensions of the standard model.
\end{abstract}
%]

\maketitle

\section{Introduction}
\label{introduction}

The Standard Model of particle interactions has passed numerous
experimental tests \cite{Hagiwara:fs} but despite the experimental
successes it is commonly believed that it is still incomplete. The
spontaneous breaking of the electroweak symmetry, for example, has
been the subject of intensive studies and different models have
been proposed to try to accommodate it in a more general and
satisfactory framework. Technicolor theories \cite{Hill:2002ap}
and supersymmetric extensions \cite{Kane:2002tr} of the Standard
Model are two relevant examples.

In \cite{Sannino:2003mt} we explored the possibility that the
relativistic Bose-Einstein phenomenon
\cite{Haber:1981ts,Kapusta:aa} is the cause of electroweak
symmetry breaking. We introduced a new global symmetry of the
Higgs field and associated a chemical potential $\mu$ with the
generators of such a new symmetry. A relevant property of the
theory was that the chemical potential induced a negative mass
squared for the Higgs field at the tree level destabilizing the
symmetric vacuum and triggering symmetry breaking. The local gauge
symmetries were broken spontaneously and the associated gauge
bosons acquired a standard mass term. We also showed that while
the properties of the massive gauge bosons at the tree level are
identical to the ones induced by the conventional Higgs mechanism,
the Higgs field itself had specific Lorentz non covariant
dispersion relations. The Bose-Einstein mechanism occurs, in fact,
in a specific frame the effects of which are felt by the other
particles in the theory via electroweak radiative corrections.
Some of these corrections have been explicitly computed in
\cite{Sannino:2003mt} and the model presented in
\cite{Sannino:2003mt} is very predictive.

\begin{figure}[bth]
\vspace{-.7cm}
 \includegraphics[width=8.5truecm, clip=true]{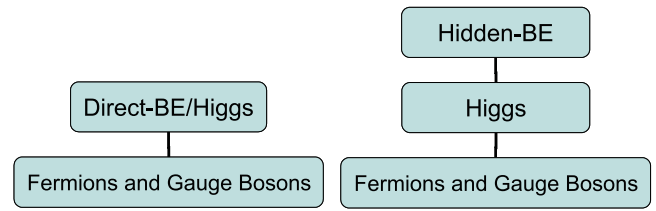}
 \caption{Schematic representation of the direct and indirect
 Bose-Einstein mechanism for triggering electroweak symmetry
 breaking.} \vspace{- .1cm}\label{Figura1}
\end{figure}

In this letter we present a new model in which a hidden
Bose-Einstein sector drives the electroweak symmetry breaking
while yielding small corrections to the standard Higgs dispersion
relations. In this way the presence of a frame and hence Lorentz
breaking corrections are suppressed with respect to the ones shown
in \cite{Sannino:2003mt}. Here the Bose-Einstein mechanism
operates on a complex scalar field neutral under all of the
Standard Model interactions. On general grounds this field
interacts with the ordinary Higgs and we use these interactions to
trigger the ordinary electroweak symmetry breaking. We investigate
some of the general consequences of the present model. Due to the
Bose-Einstein nature of the new mechanism the form of the
corrections is insensitive to its specific realization while the
mechanism is distinguishable from other sources of beyond standard
model physics. Schematically the two classes of models we imagine
are summarized in figure \ref{Figura1}. In the first one the Higgs
field feels directly the presence of a net background charge and
communicates it to the gauge bosons and fermions via higher order
corrections (left panel). In the second (the hidden case) a
singlet field with respect to the standard model quantum numbers
directly feels the effects of a net background charge. The
ordinary Higgs field weakly feels the effects of a frame via the
interactions with the singlet field. Finally the rest of the
standard model particles will be affected via higher order
corrections (right panel).
 Since the chemical potential differentiates between
time and space, and we have a scalar vacuum, all of the dispersion
relations are isotropic. Theories with condensates of vectorial
type have been studied in different realms of theoretical physics
\cite{Linde:1979pr,{Kapusta:1990qc},Ferrer:jc,Ferrer:ag,{Ambjorn:1989gb},{Kajantie:1998rz},{Sannino:2002wp}}.
It is also worth mentioning that, although in a different
framework, effects of a large lepton number on the spontaneous
gauge symmetry breaking for the electroweak theory at high
temperature relevant for the early universe have been studied
\cite{Linde:1979pr,{Ferrer:jc},{Ferrer:ag},{Bajc:1999he},{Bajc:1997ky}}.
In \cite{Mazur:2001fv} the concept of Bose-Einstein condensation
in the gravitational systems was considered.

\section{Hidden Bose-Einstein Sector}\label{due}
The Higgs sector of the Standard Model possesses, when the gauge
couplings are switched off, an $SU_L(2)\times SU_R(2)$ symmetry.
The full symmetry group is mostly easily recognized when the Higgs
doublet field is represented as a two by two matrix in the
following way:
\begin{eqnarray}
M=\frac{1}{\sqrt{2}}\left(\sigma + i\, \vec{\tau}\cdot\vec{\pi}
\right)
 \ .
\end{eqnarray}
The $SU_L(2)\times SU_R(2)$ group acts linearly on $M$ according
to:
\begin{eqnarray}
M\rightarrow g_L M g_R^{\dagger} \qquad {\rm and} \qquad g_{L/R} \in SU_{L/R}(2)\ .
\end{eqnarray}
The $SU_L(2)$ symmetry is gauged by introducing the weak gauge
bosons $W^a$ with $a=1,2,3$. The hypercharge generator is taken to
be the third generator of $SU_R(2)$. The ordinary covariant
derivative acting on the Higgs, in the present notation, is:
\begin{eqnarray}
D_{\mu}M&=&\partial_{\mu}M -i\,g\,W_{\mu}M +
i\,g^{\prime}M\,B_{\mu} \ , \nonumber \\
W_{\mu}&=&W_{\mu}^{a}\frac{\tau^{a}}{2} \ ,\quad
B_{\mu}=B_{\mu}\frac{\tau^{3}}{2} \ .
\end{eqnarray}
We now extend the standard model to contain an extra complex
bosonic field $\phi$ which is a singlet under all of the standard
model charges. In this way we also introduced an extra $U(1)$
symmetry. The most general and renormalizable potential involving
the standard model Higgs and the field $\phi$ respecting all of
the symmetries at hand is:
\begin{eqnarray}
V\left[\phi,M\right]&=&\frac{1}{2}\left(M^2_{H}-8
\hat{g}\,\left|\phi\right|^2\right){\rm
Tr}\left[M^{\dagger}M\right] + m^2|\phi|^2 \nonumber \\
&+&{\hat{\lambda}}\left|\phi\right|^4 + {\lambda} {\rm
Tr}\left[M^{\dagger}M\right]^2 \ .
\end{eqnarray}
We have assumed the potential to be minimized for a zero vacuum
expectation values of the fields and the couplings to be all
positive. With this choice is clear that if $\phi$ acquires a non
zero vacuum expectation value the ordinary Higgs also acquires a
negative mass square contribution.

We introduce a non zero background charge for the field $\phi$.
This is achieved modifying the $\phi$ kinetic term as follows
\cite{Haber:1981ts,Kapusta:aa,{Sannino:2003mt}}:
\begin{eqnarray}{\cal L}_{\rm charge}=D_{\mu}\phi^{\ast} D^{\mu}\phi\ , \label{bose}
\end{eqnarray}
with
\begin{eqnarray}
D_{\nu} \phi=\partial_{\nu}\phi - i A_{\nu}\phi \ , \qquad
A_{\nu}=\mu \left(1,\vec{0}\right) \ , \label{covariant}
\end{eqnarray}
and $\mu$ is the associated chemical potential. Substituting
(\ref{covariant}) in ({\ref{bose}}) we have:
\begin{eqnarray}
{\cal L}_{\rm charge}=\partial_{\mu}\phi^{\ast}\partial^{\mu}\phi
+i\,\mu\left(\phi^{\ast}\partial_{0}\phi -
\partial_{0}\phi^{\ast}\phi\right) + \mu^2\,|\phi|^2\ . \label{charge}\end{eqnarray}
The introduction of the chemical potential has broken Lorentz
invariance $SO(1,3)$ to $SO(3)$ while providing a negative mass
squared contribution to the $\phi$ boson. When $\mu>m$ the
spontaneous breaking of the $U(1)$ invariance is a necessity. Once
the bosonic field has acquired a vacuum expectation value and for
$8\hat{g}\,\langle |\phi| \rangle^2 > M^2_H $ the Higgs field
condenses as well.

In this model the Bose-Einstein mechanism, although indirectly,
still triggers electroweak symmetry breaking while the effects of
Lorentz breaking induced by the presence of the chemical potential
are attenuated. The latter are controlled by the new coupling
constant $\hat{g}$ as well as the ordinary higher order
electroweak corrections. Our model potential is similar in spirit
to the one used in hybrid models of inflation \cite{Linde:1993cn}.

Without any loss of generality we set $M^2_H=m^2=0$. Defining
\begin{eqnarray}
\langle |\phi| \rangle = R\, \cos\alpha \qquad \langle \sigma
\rangle = R\, \sin\alpha \ ,
 \end{eqnarray}
the potential has four extrema which are for $R=0$ and any
$\alpha$, for $R^2=\mu^2/2\hat{\lambda}$ and $\alpha=0,\pi$ or:
\begin{eqnarray}\tan^2\alpha=2\frac{\hat{g}}{\lambda}\equiv \epsilon^2 \ , \label{alpha}\end{eqnarray}
and \begin{eqnarray}
R^2=\frac{\mu^2}{2}\frac{\cos^2\alpha}{\lambda \sin^4\alpha +
\hat{\lambda}\cos^4\alpha - \hat{g}\sin^2\left(2\alpha\right)} \
.\label{Radius}
\end{eqnarray}
The respective values of the potential evaluated on the extremum
are $\langle V \rangle =0$ for $R=0$, $\langle V
\rangle=-\mu^4/(4\hat{\lambda})$ for $R\ne 0$ and $\alpha =0,\pi$
while
\begin{eqnarray} \langle V \rangle =
-\frac{\mu^4}{4\hat{\lambda}}\frac{1}{\left(1 -
\frac{4\hat{g}^2}{\hat{\lambda}\lambda}\right)} \ ,
\end{eqnarray} for the last extremum, eq.~(\ref{alpha}) and eq.~(\ref{Radius}), proving that this is also the actual ground state of the theory.
{}In the limit $\epsilon\ll 1$ we have
\begin{eqnarray}  \quad \langle |\phi| \rangle^2 =
\frac{\mu^2}{2\hat{\lambda}} + {\cal O}(\epsilon^4) \ , \qquad
\langle \sigma \rangle ^2 =
\frac{\mu^2}{2\hat{\lambda}}\,\epsilon^2+ {\cal O}(\epsilon^6)
\nonumber  \ ,
\end{eqnarray}
Assuming the new physics scale associated to $\langle |\phi|
\rangle$ to be within reach of LHC, i.e., of the order of $1-
10$~TeV while the Higgs scale is $\langle \sigma \rangle\simeq
250$~GeV we determine:
\begin{eqnarray}
\epsilon \simeq 0.25-0.025 \ .
\end{eqnarray}
All the corrections will be proportional to the fourth power of
$\epsilon$.  To explore further our model it is convenient to
adopt the unitary gauge for the Higgs field and use the
exponential map for the fields $\phi$. So we define:
\begin{eqnarray} \sigma =\langle \sigma \rangle + h \ , \quad
\phi=\frac{\left( \sqrt{2}\langle \phi \rangle + \psi
\right)}{\sqrt{2}} \,e^{i\frac{\eta}{\sqrt{2}\langle \phi
\rangle}} \ .
\end{eqnarray}
In the broken phase $h$ and $\psi$ are not mass eigenstates. These
are related to $h$ and $\psi$ via:
\begin{eqnarray}
h=\cos\theta \, \widetilde{h} - \sin\theta \, \widetilde{\psi} \
,\qquad \psi=\cos\theta\, \widetilde{\psi} + \sin\theta\,
\widetilde{h} \ .
\end{eqnarray}
The mixing angle is:
\begin{eqnarray}
\tan 2\theta
=2\sqrt{2}\frac{\lambda}{\hat{\lambda}}\,\frac{\epsilon^3}{1-2\frac{\lambda}{\hat{\lambda}}\epsilon^2
}=2\sqrt{2}\frac{\lambda}{\hat{\lambda}}\,\epsilon^3+{\cal
O}(\epsilon^5)\ .
\end{eqnarray}
The eigenvalues which, in absence of the chemical potential, one
calls the masses are:
\begin{eqnarray} m^2_{\widetilde{h}}&=&m^2_h\left[1 - 2
\left(\frac{\lambda}{\hat{\lambda}}\right)^2\,{\epsilon^6} + {\cal
O}(\epsilon^8) \right] \ , \nonumber \\
m^2_{\widetilde{\psi}}&=&m^2_{\psi}\left[1+\frac{\lambda}{\hat{\lambda}}\,\epsilon^4
+ {\cal O}(\epsilon^6)\right]\ ,
\end{eqnarray} with
\begin{eqnarray} m^2_\psi=2\mu^2 \ , \qquad m^2_h=4\mu^2\,\frac{\lambda}{\hat{\lambda}}\epsilon^2 \ .\end{eqnarray}
The Lorentz breaking term induced by the Bose-Einstein
condensation is the second term in (\ref{charge}). The new neutral
Higgs field $\widetilde{h}$ due to the mixing with $\psi$ feels
feebly but directly the presence of the net background charge. The
relevant terms in the Lagrangian are:
\begin{eqnarray}
&&\frac{1}{2}\partial_{\mu} \widetilde{h}\partial^{\mu}{\widetilde
h } + \frac{1}{2}\partial_{\mu}
\widetilde{\psi}\partial^{\mu}{\widetilde \psi
}+\frac{1}{2}\partial_{\mu}{\eta}\partial^{\mu}{\eta}-\frac{m^2_{\widetilde{h}}}{2}{\widetilde h}^2 -\frac{m^2_{\widetilde{\psi}}}{2}{\widetilde \psi}^2\nonumber \\
&&-\mu\,\cos\theta\, \left(\widetilde{\psi}\partial_0\eta -
\eta\partial_0\widetilde{\psi}\right) - \mu\,\sin\theta \,
\left(\widetilde{h}\partial_0\eta -
\eta\partial_0\widetilde{h}\right) \ . \nonumber \\
\end{eqnarray}
The propagator for $\widetilde{h}$ determined by inverting the
three by three kinetic quadratic terms is:
\begin{eqnarray}
\frac{i}{p^2-m^2_{\widetilde{h}}-4 p^2_0 \mu^2 \,\sin^2\theta
\,{\cal F}\left[p,p_0 \right]} \ , \label{propagator}
\end{eqnarray}
with
\begin{eqnarray}
{\cal F}\left[p,p_0\right]=\frac{m^2_{\widetilde{\psi}}-
p^2}{\left(m^2_{\widetilde{\psi}}-
p^2\right)\,p^2+4p_0^2\,\mu^2\,\cos^2\theta} \ .
\label{propagator2}
\end{eqnarray}
We take $\psi$ to be heavy so that the only corrections will be
the ones induced by the small mixing between ${h}$ and $\psi$. The
function (\ref{propagator2}) for large $\mu$ is ${\cal F}\approx
1/(3p^2_0-{\bf {p}}^2)$ and resembles the pole associated to the
gapless state $\eta$. If phenomenologically needed we can always
give a small mass (with respect to the scale $\mu$) to $\eta$ as
shown in \cite{Sannino:2003mt}. The rest of the standard model
particles are affected by the presence of a frame via weak
radiative corrections. These corrections can be computed as in
\cite{Sannino:2003mt}. The Higgs propagator here is more involved
than the one in \cite{Sannino:2003mt}. However the relevant point
is that the effects of the non standard dispersion relations in
the present model are depleted with respect to the ones determined
in \cite{Sannino:2003mt} due to the presence of the suppression
factor $\sin^2\theta\approx \epsilon^6$ in the $\widetilde{h}$
propagator. Indeed since $m^2_{\widetilde{h}}\approx\epsilon^2
\mu^2$ the term in the propagator bearing information of the
explicit breaking of Lorentz invariance due to the Bose-Einstein
mechanism is down by $\epsilon^4$ with respect to the mass term.
So we have a large suppression of Lorentz breaking effects induced
by the presence of a frame needed for the Bose-Einstein mechanism
to take place. Clearly also the dispersion relations of the
ordinary Higgs are modified only slightly with respect to the
standard scenario. The corrections due to the mixing with the
$\psi$ field should also be taken into account although in general
they will be further suppressed due to the assumed hierarchy
$\mu\gg m_h$.

Let us roughly estimate, using the results of
\cite{Sannino:2003mt}, the size of the corrections to some
observables. We first consider the modification of the low energy
effective theory for the electroweak theory due to the new sector.
We focus on the charged currents since the neutral currents are
affected in a similar way. The chemical potential leaves intact
the rotational subgroup of the Lorentz transformations, so the
effective Lagrangian modifies as follows \cite{Sannino:2003mt}:
\begin{eqnarray}
{\cal L}^{\rm CC }_{\rm Eff}&=&-2\sqrt{2}\,G
J_{\mu}^{+}{J^{\mu}}^{-} \Rightarrow  \nonumber \\
&-& 2\sqrt{2}\,G\,\left[{J_{\mu}}^{+}{J^{\mu}}^{-}+ \delta\,
J_{i}^{+}{J^{i}}^{-}\right] \ ,
\end{eqnarray}
where $\delta$ is a coefficient effectively measuring the
corrections due to modified dispersion relations of the gauge
bosons. Using the previous Lagrangian the decay rate for the
process $\mu\rightarrow e \bar{\nu}_{e}\nu_{\mu}$ is
\cite{Sannino:2003mt}:
\begin{eqnarray}
\Gamma[\mu\rightarrow e \bar{\nu}_{e}\nu_{\mu}]=\frac{G^2
M_{\mu}^5}{192\pi^3}\left(1 + \frac{3}{2}\delta\right) \ ,
\end{eqnarray}
where $M_{\mu}$ is the muon mass and we neglected the electron
mass. However the effects of a non zero electron mass are as in
the Standard Model case \cite{Hagiwara:fs}. The parameter $\delta$
in the model presented in \cite{Sannino:2003mt} was explicitly
computed yielding $\delta \approx 0.007$. {}For the hidden
Bose-Einstein sector this result, on general grounds, is further
suppressed by a multiplicative factor of the order $\epsilon^4$
yielding a new $\delta$ of the order of $2.7\times 10^{-5} - 2.7
\times 10^{-9}$ for $\langle |\phi| \rangle \approx 1 - 10$~TeV.

The fermion sector also bears information of the direct or
indirect nature of the underlying Bose-Einstein phenomenon. We
demonstrated for the direct case in \cite{Sannino:2003mt} that the
one loop corrections to the fermion velocities due to the exchange
of the Higgs are tiny ($\simeq 10^{-15}$ for the electron). This
is due to the smallness of the associated Yukawa's couplings.
However we have also argued that the higher order corrections due
to the modified gauge boson dispersion relations may be relevant
and estimated in this case a correction of the order of
$\delta\,g^2/4\pi$ with $g$ the weak coupling constant. In the
hidden Bose-Einstein case also the corrections to this sector are
further suppressed by a factor of the order of $\epsilon^4$ with
respect to the findings in \cite{Sannino:2003mt}.

\section{Discussion} \label{cinque}

The Bose-Einstein condensation phenomenon has been proposed as
possible physical mechanism underlying the spontaneous symmetry
breaking of cold gauge theories. In \cite{Sannino:2003mt} the
Higgs field was assumed to carry global and local symmetries and
was identified with the Bose-Einstein field. The effects of a non
zero background charge were, in this way, maximally felt in the
Higgs sector and then communicated via weak interactions to the
other standard model particles. Here we have presented a new model
in which the Higgs mechanism is triggered by a hidden
Bose-Einstein sector. The relevant feature of this model is that
the effects of modified dispersion relations for the standard
model fields are strongly suppressed with respect to the one in
\cite{Sannino:2003mt}. We have predicted the general form of the
corrections and estimated their size. The fermion sector of the
Standard Model, for example, is affected via higher order
corrections leading to the appearance of modified dispersion
relations of the type $E^2 = v_{f}^2\,p^2+m_f^2$. The deviation
with respect to the speed of light for the fermions is small when
considering the direct Bose-Einstein mechanism
\cite{Sannino:2003mt} while it is further suppressed by a factor
$\epsilon^4$ in the model presented here.

We emphasize that the form of the corrections, induced by the
modified propagators in Eqs. (\ref{propagator}) and
(\ref{propagator2}), are general and dictated solely by the
Bose-Einstein nature of our mechanism. The strength of the
coupling between the standard model fields and the Bose-Einstein
field is measured by the parameter $\epsilon$ which enters in some
of the physical observables. Experiments can be used to constrain
the possible values of $\epsilon$.

Spontaneous breaking of a gauge theory via Bose-Einstein
condensation necessarily introduces Lorentz breaking since a frame
must be specified differentiating time from space. We recall that
the issue of Lorentz breaking has recently attracted much
theoretical
\cite{Kostelecky:2002hh,Carroll:2001ws,{Amelino-Camelia:2002dx}}
and experimental attention \cite{Carroll:vb}. In the Bose-Einstein
case the underlying gravitational theory is not the cause of
Lorentz breaking which is instead due to having immersed the
theory in a background charge.

Modified dispersion relations for the standard model particles
have also cosmological consequences since now one cannot simply
assume the velocity of light as the common velocity for all of the
elementary particles. In fact different particles will have, in
general, different dispersion relations and hence speed. This is
relevant, for example, when observing neutrino which propagated
over long distances. Using the present model as guide for the
structure of the corrections experiments can test the
Bose-Einstein mechanism as possible source of electroweak symmetry
breaking.

\acknowledgments We thank P.H. Damgaard and J. Schechter for
discussions and a critical reading of the manuscript. A.D. Jackson
and K. Kainulainen are thanked for discussions. The work of F.S.
is supported by the Marie--Curie fellowship under contract
MCFI-2001-00181.

\end{document}